\documentclass[aps,prl, twocolumn]{revtex4}
\usepackage{amsmath}
\usepackage{amsfonts}
\usepackage{amssymb}
\usepackage{dcolumn}
\usepackage{bm}
\usepackage{graphicx}
\usepackage{epstopdf}
\usepackage{color}
\def \ie {{\it i.e.} }

\def\BaOsS {Ba$_2$NaOsO$_6$ }
\def\BaOs  {Ba$_2$NaOsO$_6$}
  \def \Na {$^{23}$Na }

    \def \dqS {$\delta_q$ }
   \def \dq {$\delta_q$}

\def\dq {$\delta_{\rm q}$}
\def\dqS {$\delta_{\rm q}$ }

\def\hu {$H_{\rm u}$ }
\def\hst {$H_{\rm stag}$ }
 
\begin{document}

\title{Magnetism and Local Symmetry Breaking  in  a Mott Insulator with Strong Spin Orbit Interactions}

\author{ L. Lu$^{1}$, M. Song$^{1}$, W. Liu$^{1}$,  A. P. Reyes$^{2}$, P. Kuhns$^{2}$,  H. O. Lee$^{3,4}$, I. R. Fisher$^{3,4}$,  
  and  V. F. Mitrovi{\'c}$^{1, \dagger}$}
  
\address{$^{1}$Department of Physics, Brown University, Providence, RI 02912, U.S.A.\\
$^{2}$National High Magnetic Field Laboratory, Tallahassee, FL 32310, USA\\
$^{3}$Department of Applied Physics and Geballe Laboratory for Advanced Materials, Stanford University, California 94305, USA\\
$^{4}$Stanford Institute for Materials and Energy Sciences, SLAC National Accelerator Laboratory, 2575 Sand Hill Road, Menlo Park, California 94025, USA }
\date{\today}  

\begin{abstract}

{Study of the combined effects of strong electronic correlations with spin-orbit coupling (SOC) represents a central issue in quantum materials research. Predicting emergent properties represents a huge theoretical problem since the presence of SOC implies that the spin is not a good quantum number. Existing theories propose the emergence of a multitude of exotic quantum phases, distinguishable by either local point symmetry breaking or local spin expectation values,  even in materials with simple cubic crystal structure such as Ba$_2$NaOsO$_6$. Experimental tests of such   theories by local probes are highly sought for.  Here,  we report on  local  measurements designed to concurrently probe spin and orbital/lattice degrees of freedom of Ba$_2$NaOsO$_6$. 
We find that  a novel canted ferromagnetic phase which is  preceded by local point symmetry breaking is stabilized at low temperatures, as predicted by quantum  theories involving  multipolar spin interactions.}
   \vspace*{-0.5cm}
\end{abstract}

\maketitle

{

{\bf Introduction}

  Magnetic Mott insulators with strong spin-orbit coupling (SOC) represent an intriguing class of materials   where various exotic quantum phases,   that include spin liquid, multipolar charge order, topological insulator and semimetal, Weyl semimetal, and Axion insulator, are predicted to emerge \cite{Chen09, Jackeli09, ChenBalents10, ChenBalents11, Pesin10, Radic12, Cole12, Reuther11, KrempaRev14, Nussinov15}. SOC is a relativistic effect  that in the  strong regime leads to  local entanglement of spin and orbital degrees of freedom. This entanglement  results in drastically different physics  than in cases of weak  SOC.     The associated emergent phenomena are particularly rich in metal oxides containing $5d$ transition-metal ions owing to the comparable magnitude of both strong electron correlations and SOC. 
Particularly interesting is the case of materials with a  double perovskites structure \cite{Aharen10, Erickson07, Yamamura2006, Wiebe03, Wiebe02}, for which it has been proposed that  partial  lifting of degeneracy of the total angular momentum eigenstates induces a highly nontrivial multipolar exchange interactions \cite{ChenBalents10,ChenBalents11}. 
These peculiar interactions promote quantum fluctuations and thus generate novel quantum states impossible without strong SOC \cite{ChenBalents10,ChenBalents11,Dodds11}.  Such states include   an unconventional antiferromagnet with the dominant magnetic octupole and quadrupole moments,  an unusual noncollinear ferromagnet   with a doubled unit cell and magnetization along the [110] axis,  and  biaxial spin nematic  phase with quadrupolar order    with  preserved  time-reversal symmetry 
stabilized in a broad intermediate temperature range  above any magnetic ordering temperature.      A key feature of these many-body quantum models is 
  that significant interactions are  fourth and sixth order in the effective spins, due to    strongly orbital-dependent exchange.

  A representative material in this class is \BaOs,    a double perovskite with Na and Os ions inhabiting alternate cation ÒBÓ sites, which for an undistorted structure has a  face-centered-cubic lattice,   as shown in \mbox{Fig. \ref{Fig1}c}. 
   Thermodynamic  and reflectivity measurements  characterize this material 
     as a $5d^1$ ferromagnetic (FM) Mott insulator with a moderate
ordered moment $\sim 0.2\, \mu_{\rm B}$ per formula unit  and \mbox{$T_{\rm c} \sim  6.3$ K}     \cite{Stitzer2002, Erickson07}. 
    This relatively small value of the ordered moment was confirmed in $\mu$SR measurements \cite{Steele11}.
     Taking SOC into account,  the anticipated ground state for a perfectly cubic point symmetry  is  $J_{\rm {eff}}={3 \over 2}$.  
Yet, the magnetic entropy removed at FM transition  is only $R  \ln2$ \cite{Erickson07}.   
 Though,  the most unusual observation is that  the FM state   easy axis is in the [110] direction, as this does not occur in standard Landau theory for ferromagnetism in a cubic symmetry \cite{ChenBalents10}. This uncommon magnetism can either be explained by the  density functional theory (DFT) electronic structure calculations,   that include 
 effects of electron correlation,  a strong SOC, and anisotropic exchange interaction \cite{Pickett15}; or, by  quantum models  including multipolar exchange interactions arising from strong SOC  \cite{ChenBalents10, Balents14}.
Moreover,  quantum models   identify   the  quadrupolar/orbital ordering, as a driving mechanism for the FM phase  that develops atop it \cite{ChenBalents10, Balents14}.  
This quadrupolar order is characterized by the orbital polarization that is distinct on the two sublattices. As such polarizations  cannot be time-reversal conjugates,    when magnetism onsets, a net ferromagnetic moment results.   
Moreover, because this anticipated quadrupolar ordering, that manifests in  a breaking of the local cubic symmetry, onsets at  higher temperature than magnetic order,  many-body  quantum based models also account for the missing entropy at the FM transition \cite{Erickson07, Balents14} (see Supplementary Discussion 3).  
  Experimental confirmation of the microscopic quantum models requires the observation of two effects. These comprise a structural change, that precedes magnetic order, associated with the quadrupolar ordering   and   local spin expectation values, that differ from the average ones.  
 Here, we report   the first  observation of such effects. Namely, we observe exotic  local spin expectation values along with the 
 structural changes,  and  infer the exact microscopic  nature of the FM state and lattice distortions. 
 We show that the FM state is in fact a type of canted  ferromagnet  with two sub-lattice magnetization, and that cubic symmetry breaking  occurs at a temperature above the N{\'e}el temperature  and it involves deformation of oxygen octahedra  presumably reflecting a complicated pattern of staggered orbital order. 
  In light of our NMR findings, we compiled  the   phase diagram sketched in \mbox{Fig. \ref{Fig1}b}.
 Our results are in startlingly good agreement with recent theoretical predictions based on quantum models \cite{ChenBalents10, Balents14}. Thus,   our findings establish that such quantum models represent an appropriate theoretical framework for predicting emergent properties in  materials with both strong correlations and SOC, in general.

%
\begin{figure}[t]
\begin{minipage}{0.98\hsize}
 \centerline{\includegraphics[scale=0.45]{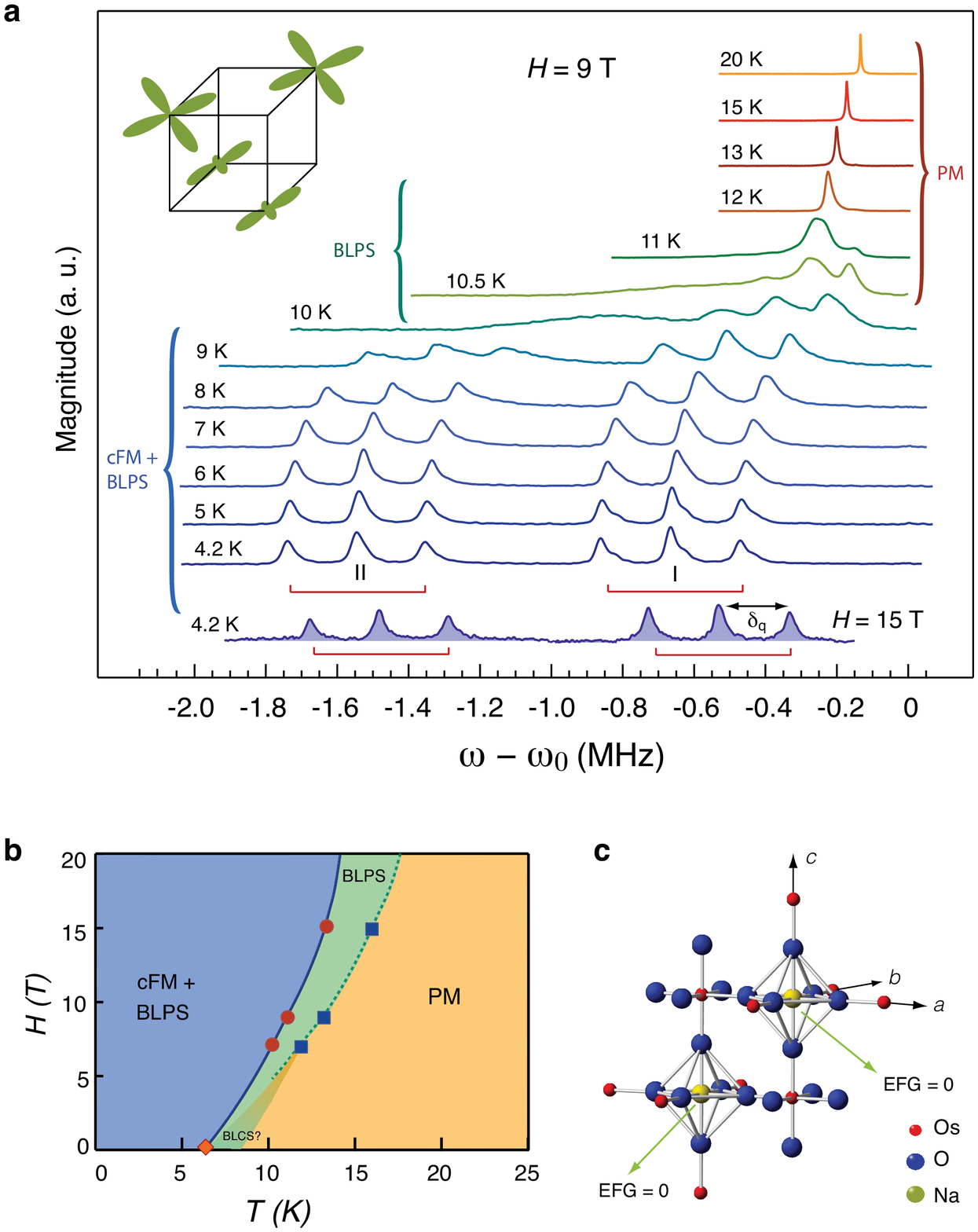}}  
\begin{minipage}{.98\hsize}
 \vspace*{-0.3cm}
\caption[]{\label{Fig1} 
{\bf Phase diagram of  \BaOsS deduced from the  NMR spectra.} 
}
 \vspace*{-0.2cm}
\end{minipage}
\end{minipage}
\end{figure}
%

\vspace*{0.2cm}
{\bf Results}

 {\bf NMR Spectra.} 
In a lattice with  cubic symmetry only  one Na NMR site is present, which results in a  narrow single peak spectrum in the insulating paramagnetic (PM) state.  
 Our  main discovery is that, on reducing temperature, the Na NMR line undergoes a complex modification, as evident  in \mbox{Fig. \ref{Fig1}a}, reflecting changes in local electron spin susceptibility and electric field gradient (EFG). 
 That is, we deduce that this modification reflects a splitting of Na into two sites sensing different hyperfine fields due to electronic spins and  breaking of the  local cubic point symmetry with  both sites  sensing the same quadrupole frequency, related to the EFG.

   To establish the sensitivity of our measurements to putative lattice distortions, orbital order, and magnetism, 
   we first inspect the temperature $(T)$ dependence of   $^{23}$Na NMR spectra in \BaOs.  
 These   $^{23}$Na  
 spectra reveal the distribution of    the hyperfine fields and the electronic charge and are thus a sensitive probe of both the electronic spin polarization (local magnetism) and  charge distribution (orbital order and lattice symmetry).   Temperature evolution of   $^{23}$Na NMR spectra is plotted in  \mbox{Fig. \ref{Fig1}a}. 
For  $T > 12 \, {\rm K}$,  the spectrum consists of a single narrow NMR line,    evidencing    the PM state.  
Below \mbox{13 K}, the NMR line broadens and splits into multiple peaks indicating onset of significant changes in the 
 the local symmetry, thereby producing EFG, that is asymmetric (non-cubic) charge distribution. 
  Below 10 K, the  \Na spectra clearly split into 6 peaks, that is two sets of triplet lines,   labeled  as I and II in \mbox{Fig. \ref{Fig1}a}, that are well separated in frequency.  
  
   {\bf Two magnetic sites.} 
  Broadly  speaking,  the emergence  of these two sets of  triplets  indicates appearance of  two distinct magnetic sites, \ie two nuclear sites that 
  sense two different local fields,  in the lattice. 
  In a field of 9 T, the  transition to  the LRO  state onsets  in the vicinity of 10 K, which is 
 significantly higher than the transition temperature observed in low field  thermodynamic measurements in \cite{Erickson07}. Our data measured at the different applied magnetic fields indicate that the transition temperature increases with the increasing field, confirming  the magnetic nature of the transition (see Supplementary Note 2).  
The line shape and the  fact that the two  sets of triplet lines, I and II,  are well separated in frequency  implies that the  low temperature  LRO  magnetic state is commensurate.  Furthermore,    both sets of lines are shifted to frequencies below that of spectra in the   PM state.
This demonstrates that the net local magnetic fields on both Na sites  are of the same sign indicating that the LRO   order is likely ferromagnetic.  

   {\bf Point symmetry breaking and EFG.} 
Besides the  splitting into sets I and II, that reflects the appearance of two distinct magnetic sites in the low $T$ phase,  
 we observe additional splitting of each set of the  spectral lines into three peaks. 
  Moreover,  as visible in \mbox{Fig. \ref{Fig1}a,}   the additional splitting is discernible  at temperatures higher than that for the onset of LRO state, apparent in  emergence of sets I and II.
 This  splitting, labeled as \dqS in \mbox{Fig. \ref{Fig1}a}, originates from quadrupole interaction,  implying changes in local charge distribution induced by modifications of     electronic orbitals and/or local lattice symmetry.
\onecolumngrid
\begin{center}
  %
 %
\begin{figure}[t]
  \vspace*{-0.0cm}
\begin{minipage}{0.98\hsize}
 \centerline{\includegraphics[scale=0.75]{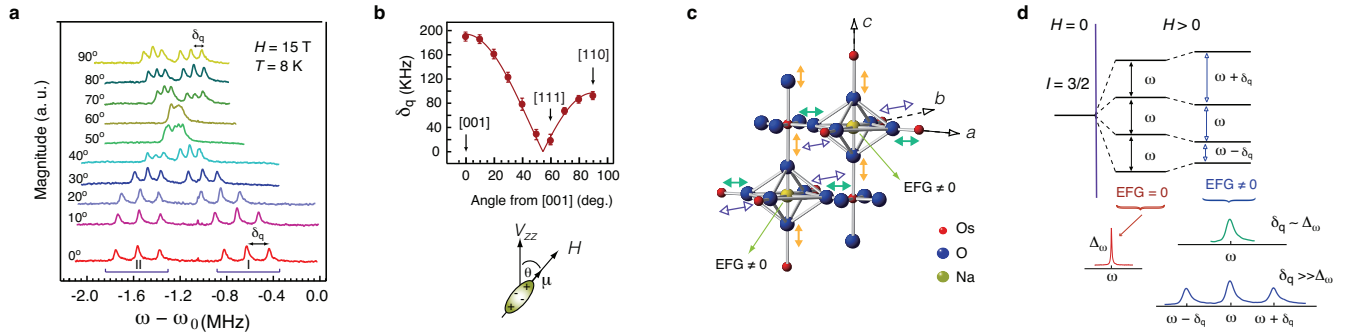}} 
\begin{minipage}{.98\hsize}
 \vspace*{-0.0cm}
\caption[]{\label{Fig2} 
{\bf Local cubic symmetry breaking in the ordered phase.}  
}
 \vspace*{-0.3cm}
\end{minipage}
\end{minipage}
\end{figure}
%
\end{center}
  \vspace*{-0.90cm}
 \twocolumngrid
%
 %
%
  
  For nuclear sites with spin  $I  >   1/2$, such as $^{23}$Na with $I=3/2$, and non-zero   EFG,   quadrupole interaction between nuclear spin and EFG splits otherwise single NMR line to $2I$ lines,  as illustrated in \mbox{Fig. \ref{Fig2}d}. 
 Thus, $^{23}$Na spectral line splits in three in the presence of non-zero EFG.  
 Nonetheless, for small finite values of the EFG  three peaks are not necessarily discernible,  in which case  significant line broadening can only be observed, as depicted  in \mbox{Fig. \ref{Fig2}d}. 
 At  sites with  cubic point   symmetry the EFG is zero  
 (see Supplementary Note 5), as is the case for Na nuclei in the high temperature PM phase.   Therefore, the observed line broadening and subsequent splitting of the Na spectra into triplets, in the magnetically    ordered phase, indicates breaking of the  cubic point symmetry, caused by local distortions of 
 electronic charge distribution.  These distortions,   marking the broken local point symmetry (BLPS) phase, occur above  the transition into the magnetic state,  as depicted in \mbox{Fig. \ref{Fig1}b}. 
  To confirm this finding,  we  measured low $T$ spectra as a function of strength and   orientation of the applied magnetic field,  as we describe next. 
  
  Specifically, we examine \dq, the average separation between two adjacent quadrupolar satellite lines, as a function of strength $(H)$ and   orientation $({\mathbf H})$ of 
 the applied magnetic field. The size of  \dqS is proportional to the magnitude of the EFG and the square of the spin operator $(\hat I_{\rm z})$ 
 projected along the principal axis of the EFG $(V_{\rm zz})$  (see Supplementary Note 5). In strong applied field, as is the case in our experiment,  the size of \dqS is controlled by 
  the projection of   $\hat I_{\rm z}^{2}$  along ${\mathbf H}$, proportional to the $\cos^{2}(\theta)$ of the angle  $ \theta$ between ${\mathbf H}$ and $V_{\rm zz}$,  as shown in  \mbox{Fig. \ref{Fig2}b}. Evidently, \dqS is at its maximum for ${\mathbf H}$  applied in the direction of $V_{\rm zz}$, which in our experiment   corresponds to [001] direction. 
 Therefore,  if the splitting \dqS originates from quadrupole interactions,  \dqS should remain constant as strength of the the applied field is changed and  should follow  $|(3   \cos^{2} \theta -1)/2 |$   function of $\theta$ as its orientation is varied \cite{AbragamBook}.  
We compare  \dqS    in 
 fields ranging from \mbox{7 T} to \mbox{15 T} at \mbox{4 K}, deep in the LRO phase (see \mbox{Fig. \ref{Fig1}a}). 
 Comparison reveals that  \dqS  varies by less than 2 \%, which is of the order of the error bars. 
Furthermore, we measured the spectra at 15 T and 8 K as a function of the of the angle $(\theta)$ between ${\mathbf H}$ 
  and [001] crystalline axis as plotted in   \mbox{Fig. \ref{Fig2}a}. 
  The angle dependence of the splitting \dqS is displayed in  \mbox{Fig. \ref{Fig2}b}.  It indeed follows the exact functional dependence expected for the \dqS originating from quadrupole interactions. 
    Both the observed insensitivity of \dqS   to the strength of the magnetic field and  its  dependence on $\theta$ indicate that  \dqS  splitting originates from   structural/orbital distortions for which   the principal axes of the EFG coincide with those of the crystal (see Methods).   
    Moreover, the insensitivity of \dqS   to the strength of the magnetic field     
    rules out the possibility that the detected distortions  originate from trivial magnetostriction effects on the crystal. 
    Our finding, that structural distortion  is  present in the   LRO phase, is in contrast to the predictions made by the first-principles 
DFT calculation  \cite{Xiang07, Pickett15}. 
This is important in so far that it clearly shows that  quantum models based on 
 complex multipolar  interaction generating high-order spin exchange   is consistent with 
  the observed nature of emergent phases in Mott insulators with the strong SOC \cite{ChenBalents10, Balents14}. 
  
    {\bf Lattice distortions and orbital order.} 
To resolve  the microscopic nature of the observed lattice distortions and determine their magnitude, we  performed detailed numerical calculations of the EFG,
and thus \dq, based on the  point charge approximation \cite{AbragamBook} (see Supplementary Note 5). We found that our observation, revealing equal \dqS on two magnetically inequivalent Na sites, can be best      explained by a  scenario involving  
 distortions of the O$^{2-}$ octahedra, surrounding Na$^{+}$ ions as depicted in  \mbox{Fig. \ref{Fig2}c}. In this scenario,   
 one structurally distinct Na site in non-cubic environment is generated.  
As it results from our calculations, distortions that generate  orthorhombic  local symmetry at the Na site are required to account for both the amplitude of the detected splitting ($\delta_{\rm q} \approx 190\, {\rm kHz}$) and its dependence on the field orientation. Thus, in the ordered phase, our  observations are explained by the  orthorhombic  distortions   that comprise  of dominant deformations along [001] and one of the crystalline axis in the (110) plane.  
Even though we find several possible distortions which can induce the observed splitting, we emphasize that they all involve a symmetry-lowering transition to an orthorhombic point symmetry in the LRO phase.  
Considering the observed amplitude of  \dq, we   deduce that  a typical  magnitude of the distortion  along any particular direction in the LRO  phase does not exceed \mbox{0.8 \%} of the  respective lattice constant.  
 Above $T_{c}$, in the BLPS phase, the width of the NMR spectra allows us to place an upper limit on distortions. We infer that the limit equals to  0.02\% of the  respective lattice constant, as any deformations that exceed this value would cause visible  splitting of the NMR spectra in the PM state. 
 
   In  the PM state,  BLPS phase is characterized by 
 significant  broadening of the NMR spectra. This broadening grows rapidly on decreasing temperature towards $T_{\rm c}$.   
 The angle dependence of the broadening does not coincide with either  that of the internal uniform or staggered fields, indicating that the broadening predominantly originates  from lattice distortions. Because in the BLPS phase we do not observe well defined splitting, but  rather convoluted broadening, the exact 
 dependence of \dqS on the field orientation is unknown.   Consequently, dominant tetragonal distortions along [001] direction can in principle  account for the line broadening in the PM phase. Thus, the BLPS phase  can be viewed as the PM phase in which the cubic point symmetry is broken by either dominant tetragonal deformations of the oxygen tetrahedra along [001] direction or orthorhombic distortions, as is the case in the LRO phase (see Supplementary Note 5 and Supplementary Discussion 1).  Hence, it is possible that the solid line,  indicating $T_{\rm c}$ into LRO magnetic phase,  in  \mbox{Fig. \ref{Fig1}b}   denotes  symmetry lowering transition from  tetragonal-to-orthorhombic phase as well. 

     {\bf  LRO magnetic state.} 
  We  emphasize that   detected lattice distortions 
    lead to only one  magnetically distinct Na site. 
   Thus,  we conclude that the observed  magnetically    distinct  Na  sites (labeled as I and II in  \mbox{Fig. \ref{Fig1}}) must originate  from the novel type of 
    magnetism and not from lattice distortions.
  In order to deduce the microscopic nature of the LRO magnetism,  we next discuss the   temperature and field evolution of the 
    local fields.  We used the NMR shift data to infer the local uniform $(H_{\rm u} = {1 \over 2} \left[ \langle H_{\rm I} \rangle +   \langle H_{\rm II} \rangle \right ])$ and staggered $(H_{\rm stag}= {1 \over 2} \left[ \langle H_{\rm I} \rangle -   \langle H_{\rm II} \rangle \right ] )$ fields, where   
  the average is taken over the triplet I and II, as denoted (see Supplementary Notes 1 and 2 ).       
    We note that \hu corresponds to the local field as determined by the first moment of the entire spectra. 
Temperature evolution of these local fields is displayed in \mbox{Fig. \ref{Fig3}}.  Well below $T_{\rm c}$, \hu increases with increasing $H$, while  \hst (observable only by local probes) remains constant. Interestingly, both \hu and \hst are of the same order of magnitude. Presence of  \hst  implies that the LRO state contains two-sublattice magnetization with significant antiparallel components.  
  Such  magnetization  naturally accounts for the appearance of two magnetically inequivalent Na sites, that is the appearance of distinct
local fields $\langle H_{\rm I} \rangle$ and $\langle H_{\rm II} \rangle$.
   In   \mbox{Fig. \ref{Fig4}a}, the internal field at the Na site in one plane  consists of a sum of  the four Os  moments on the same layer, and thus equal magnetic moments from one sub-lattice  labeled A,  and 
    two Os above/below in neighboring layers with magnetic moment pointing in a different direction (sub-lattice  labeled B), and thus producing different local field than A moments at the Na site. 
     Na nuclei in the next plane will then sense four type B and two type A Os moments.  This generates two sets of inequivalent Na sites and  causes the magnetic splitting in spectrum between  triplet I and II, as  two types of moments induce different local fields at the Na site.     
  %
%
 %
\begin{figure}[b]
  \vspace*{ -0.3cm}
\begin{minipage}{0.98\hsize}
 \centerline{\includegraphics[scale=0.80]{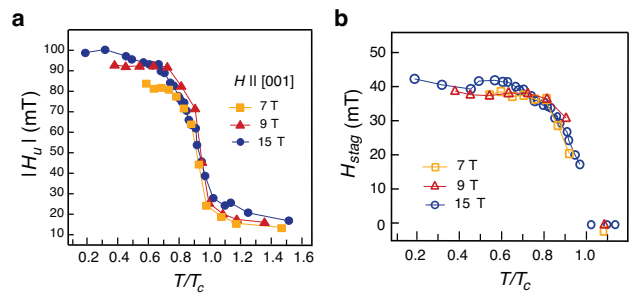}} 
\begin{minipage}{.98\hsize}
 \vspace*{-0.3cm}
\caption[]{\label{Fig3} 
{\bf Temperature dependence of the uniform and staggered fields.}  
  }
 \vspace*{-0.5cm}
\end{minipage}
\end{minipage}
\end{figure}
%

Next, we inspect   the   local fields as 
  the applied magnetic field was rotated in the $(1 \bar1 0)$ plane of the crystal, as illustrated in \mbox{Fig. \ref{Fig4}a}. 
This is essential for ensuring that we understand what components of anisotropic magnetic susceptibility are being measured.   
  For $H \| [001]$,  $H_{\rm stag}$, as well as $ \langle H_{\rm II} \rangle $, reaches its maximum value, while both  \hu and $ \langle H_{\rm I} \rangle $ are at their minimum. In principle,    \hu should scale as bulk magnetization $(M)$. As evident in \mbox{Fig. \ref{Fig4}c},  this is not the case here. This finding reveals that 
  despite the fact that the net magnetization is aligned with the [110] axis,  the local  fields, \ie spin expectation values, are not.  This very fact was predicted to arise as a direct consequence of lattice distortions driven by complex interactions in this class of materials \cite{ChenBalents10}.

   %
\begin{figure}[t]
\begin{minipage}{0.98\hsize}
 \centerline{\includegraphics[scale=0.50]{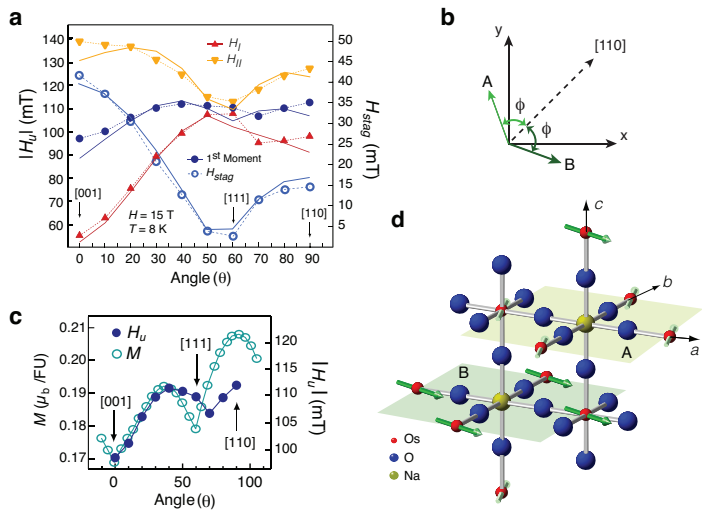}} 
\begin{minipage}{.99\hsize}
 \vspace*{-0.2cm}
\caption[]{\label{Fig4} 
{\bf  Resolution of the spin orientation in the ordered phase. }
   }
 \vspace*{-0.6cm}
\end{minipage}
\end{minipage}
\end{figure}
%
%

 The angular dependence of the internal fields is used to deduce the exact spin orientation in the LRO phase by calculating   the  local \hu and \hst at the Na site for a given spin orientation. The local field consists of a contributions from  electronic spins at six  nearest-neighbor Os  ions mediated via anisotropic hyperfine interaction (Methods). 
 By performing full lattice sum,  we calculate the  local fields at the Na site as the direction of the applied field is rotated in   $(1 \bar1 0)$ plane of the crystal. 
 We find that the model that best describes our observations,  as illustrated by the solid lines in  \mbox{Fig. \ref{Fig4}a}, is a two-sublattice  canted FM model, recently proposed in \cite{Balents14} and depicted in  \mbox{Fig. \ref{Fig4}b}. 
This model consists  of two inequivalent sub-lattices with moments in each layer in the XY plane parallel to each other, forming FM order, while moments in the neighboring layers point to a different direction.  Specifically,  moments in   two adjacent layers  are symmetric about [110] axis,   that is   they form an angle $\pm \phi$, with [110] axis, as depicted in \mbox{Fig. \ref{Fig4}b and d}.  As  direction of the applied field is varied spin-plane follows the direction of ${\mathbf H}$ while spins  remain staggered about [110] axis.  Moments arranged in this fashion   induce  an uniform field in [110] direction,   providing an overall shift to the NMR spectrum, and form a staggered pattern in the direction perpendicular to  [110]. Thus, for ${\mathbf H} || [001]$ such spin arrangement generates  lowest \hu and largest $H_{\rm stag}$, as observed, when the canting angle  $\phi$ exceeds $45^\circ$. The exact value of the canting angle is determined by the fitting procedure described below.
 The curves in \mbox{Fig. \ref{Fig4}A}, are fits to the data using the 
    simulated  local fields with relative strength of the off-diagonal terms of the  hyperfine coupling tensor, $\mathbb{A}$, $(A_{ij}/A_{ii})$,  magnitude of the local Os moments, and the canting angle $\phi$ between the spins on two different  sub-lattices, as  fitting parameters. 
Constraining the diagonal terms of $\mathbb{A}$ to be close to those found in the PM state,  
we find  the  moment of $\mu \approx 0.6 \, \mu_{\rm B}$ and  $\phi \approx 67^{\circ}$. 
 Using the  deduced value of  $\phi$ and  formalism in \cite{Balents14}, we estimate   the ratio of in-plane to intra-plane coupling constant to be $\approx 4$ (see Supplementary Note 4).  
 The value of the moment is in agreement with the effective moment  
deduced from the fit to a Curie-Weiss law in the PM  state  in \cite{Erickson07}. Large canting angle $\phi$ explains the  smaller moment detected in the FM state in bulk measurements  \cite{Erickson07, Steele11} due to partial cancelation of nonparallel magnetic moments.  
We emphasize that the symmetry of the inferred $\mathbb{A}$ tensor   reflects neither local tetragonal nor orthorhombic symmetry of the distorted O octahedra  (see Supplementary Notes 3 and 4). This indicates  that spin-spin interactions are  highly anisotropic. 

\vspace*{0.2cm}
  {\bf Discussion}

We have performed microscopic measurements on a model system of Mott insulator with strong SOC, \BaOs. 
Our static NMR measurements   reveal that the local cubic symmetry breaking, induced by  deformation of the oxygen octahedra, precedes the formation of the LRO magnetism.  
  We establish that LRO state is the exotic canted two-sublattice FM state, believed to be driven by the staggered quadrupolar order \cite{ChenBalents10}. 
  This   is the first direct detection of such a  complex quantum state with the distinct   local 
 spin expectation values.  
 Our observation of both the local cubic symmetry breaking and appearance of  two-sublattice  exotic FM phase is in line with theoretical predictions based on quantum models with  multipolar magnetic interactions.  
The fact that  spin-spin interactions are indeed mediated by  complex multipolar interactions, as suggested in \cite{ChenBalents10}, 
 is further confirmed by our finding that the symmetry of the inferred $\mathbb{A}$ tensor    does not reflect any local   symmetry of the distorted O octahedra. 
Moreover, it is proposed that two-sublattice magnetic structure is the  very manifestation of staggered quadrupolar order  and that this ordering drives the formation of LRO magnetism \cite{ChenBalents10, KrempaRev14}. Thus, our finding that LRO phase is a  two-sublattice  canted FM implies   that broken cubic symmetry phase is a  staggered quadrupolarly ordered phase with distinct orbital polarization on two-subattices. In summary, our findings clearly demonstrate that  microscopic quantum models with  multipolar magnetic interactions are   an appropriate theoretical framework for predicting emergent quantum phases in  Mott insulators with the strong SOC.  \\

 Lastly,  presented unique direct observation of both local cubic symmetry breaking and exotic long range ordered (LRO)  magnetic state  
   is the confirmation of the theoretical proposal that  
   the combination of the unusual multipolar interactions, generic for   spin-orbitally entangled effective spins, and/or structural transitions  or quadrupolar order  can lead   to a highly frustrated quantum regime even  for  systems with  spin greater than  $S = 1/2$     \cite{Balents14}. Thus, our work illustrates  that such complex quantum states might be found in other   frustrated  materials with both strong correlations and SOC \cite{Cook15, Romhanyi16}. \\

          {\bf   METHODS} 
          
          \vspace*{0.2cm}
{\bf  NMR methods.} The measurements were done at Brown University for magnetic field up to 9 T and at the NHMFL in Tallahassee, FL    
at higher fields. In both laboratories high homogeneity superconducting magnets were used. 
 The temperature control was provided by $^4$He variable temperature  insert. 
The NMR data were recorded using a state-of-the-art laboratory-made NMR spectrometer.
The spectra were obtained, at each given value of the applied field, from the sum of spin-echo Fourier transforms recorded at constant frequency intervals. 
We used a standard spin echo sequence $(\pi/2-\tau-\pi)$. Shape of the spectra presented in the manuscript  
are independent of the duration of time interval $\tau$.  Since  nuclear spin $I$ of $^{23}$Na equals to $3/2$ and at low temperatures,  both Na sites (I and II) are in non-cubic environments, three distinct quadrupolar satellite lines are observed per site \cite{AbragamBook}. 
The shift was obtained from the frequency of the first moment of spectral distribution of set of triplet lines using a gyromagnetic ratio of  $^{23}\gamma$ = 11.2625 MHz/T. The same gyromagnetic ratio was used for all frequency to field scale conversions. \\

   {\bf   Sample.}       High quality single crystal of \BaOsS  with a truncated octahedral morphology were grown from a molten hydroxide flux, as described elsewhere  \cite{Stitzer2002, Erickson07}. Crystal quality was checked by x-ray diffraction, using a Bruker Smart Apex CCD diffractometer, which indicated that the room temperature structure belongs to the  $Fm\bar 3m$ space group  \cite{Erickson07}. NMR measurements were performed for a single crystal with a volume of approximately 1 mm$^3$. The quality of the sample was confirmed by the sharpness of
    $^{23}$Na  NMR spectra both in the high temperature paramagnetic state and low temperature quadrupolar split spectra.  
  
  The sample was both zero-field and field-cooled. We did not   detect any influence of the sampleÕs cooling history on the NMR spectra.  
 Nevertheless for consistency, all results presented in the paper were obtain in field-cooled conditions. 
The sample was mounted to
one of the crystal faces and rotated with respect to the applied field about an axis using a single axis goniometer. The rotation angle, for applied fields below 9 T, was inferred from the signal of two perpendicularly positioned Hall sensors. In addition, to ensure that data was taken with no external pressure applied, the mounted sample was placed in a solenoid coil with cross sectional area significantly larger than that of the sample. In this way, no  pressure   is exerted on the sample as coil contracts   on  cooling. \\

 {\bf Transition Temperature:}
 Transition temperature $(T_{c})$ from paramagnetic (PM) to low temperature ferromagnetic (FM) state was determined   by examining the NMR shift and spectral line  shapes, as described in detail in Supplementary  Note 2. 
 Onset temperature for breaking of local cubic symmetry, shown in \mbox{Fig. \ref{Fig1}b}, was identified as  temperature below which the second moment of the NMR spectral line, measuring the spectral width, 
   increases  notably as compared to that  in  high temperature PM phase. We point out that this temperature does not necessarily correspond to the true onset temperature for orbital ordering, which could be   undetectable in our experiment due to the subtlety of the effect.  \\
   
   {\bf Quadrupolar Interaction and EFG:}
   In the simplest case of a field with axial symmetry,  interaction between  $eq$, the electric field gradient (EFG),  and the nucleus, with spin $I$ and the quadrupole moment $Q$,  is described by the  Quadrupole Hamiltonian,
$ \mathcal{H}_{\rm Q} = \frac{(eQ)(eq)}{4I(2I-1)}[3I_{\rm z}^2-I(I+1)]$. 
For nuclear spin $I =3/2$, as is the case of $^{23}$Na, the energy eigenstates of $\mathcal{H}_{\rm Q}$ are given by,
$E =   \frac{(eQ)(eq)}{4I(2I-1)}\, [3m^2-I(I+1)]$. Than, the frequencies   between different quadrupole satellite transitions equal,
\begin{equation}
\begin{split}
\omega_{\rm m \rightarrow m-1}  = \,  \, &    \frac{(eQ)(eq)} {h \, 4I(2I-1)} \,[3(2m-1)]  = \\
  \frac{1}{2},   \quad &{\rm for} \quad |+ 3/2 \rangle  \rightarrow  |+1/2 \rangle\\
0,   \quad &{\rm for} \quad |+ 1/2 \rangle  \rightarrow  |-1/2 \rangle\\
 -\frac{ 1}{2},\quad &{\rm for} \quad |- 1/2 \rangle  \rightarrow  |-3/2 \rangle\, .\\
\end{split}
\end{equation}

Therefore, in a magnetic field applied along the principal axis of the EFG only 3 NMR lines (transitions) will be observed with equal splitting $\delta_{q}$. In this case, the quadrupole splitting $\delta_{q}$ between different quadrupole satellites is simply given by 
$
\delta_{\rm q} = \frac{1}{2 h} (eQ)(eq) = \frac{1}{2h} \text{(Quadrupole moment)} \times \text{(EFG)}. 
$
In our experiment equal splitting is observed between quadrupole  satellites lines plotted in \mbox{Fig. \ref{Fig1}a} for ${\mathbf H} \| [001]$ indicating that the principal axis of the EFG must be along ${\mathbf H}$, that is along the axis of the crystal. Further, we can estimate the value of the EFG using experimentally determined value of the splitting.

For anisotropic charge distributions, quadrupole Hamiltonian expressed in the coordinate system define by the principal axes of the EFG, is given by
\begin{equation}
\begin{split}
\label{ham}
\mathcal{H}_{\rm Q}(x,y) = \frac{eQV_{zz}}{4I(2I-1)} \left[(3\hat{I}^2_{z}-\hat{I}^2)+\eta(\hat{I}^2_{x}-\hat{I}^2_{y}) \right ],
\end{split}
\end{equation}
where  $\eta \equiv\left |  {V_{xx}-V_{yy}} \right |/{V_{zz}}$ is asymmetry parameter  and  V$_{\rm xx}$, V$_{\rm yy}$, and V$_{\rm zz}$ are diagonal components of the EFG.
Here,  $V_{\rm zz}$ is defined as the principle component of the EFG and  $|V_{\rm xx}|< |V_{\rm yy}|< |V_{\rm zz}|$, by convention.    
In this case, the splitting is given by,
$
\delta_{\rm q} = \frac{(eQ)( V_{\rm zz})}{2 h}   \, \left ( 1 +   \frac{\eta^{2}}{3} \right )^{{1/2}} . 
$
Thus, the value of  $\delta_{\rm q}$ is dictated by both $ V_{\rm zz}$ and anisotropy parameter. In the high field limit, when $\mathcal{H}_{\rm Q}$ is a perturbation to the dominant Zeeman term, the angular dependence of the splitting is given by
\begin{equation}
\delta_{\rm q} = \frac{\Delta_{\rm q}}{2} (3\cos^2\theta - 1 + \eta \sin^2\theta \cos2\phi),
\label{qsplit}
\end{equation}
where   $\theta$ is the angle  between the applied field $\mathbf H$ and $V_{\rm zz}$.    
As in the case of axially symmetric EFG, in the coordinate system defined by the principal axes of the EFG only 3 NMR lines (transitions) will be observed with equal splitting $\delta_{\rm q}$ between them. When $\mathbf H$ is rotated in such   coordinate system only  3 NMR lines are observed and the magnitude of the splitting between these lines depends only on  angle $\theta$. 
The fact that we observe no more than 3 lines  per set (I or II) regardless of the angle between $\mathbf H$ and [001] crystalline axis, as   shown in  \mbox{Fig. 2B},  indicates that 
$\mathbf H$ was rotated in the coordinate system defined by the principal axes of the EFG. Therefore, the principal axes of the EFG  must coincide with those of the crystal. 
 
 In a material with cubic symmetry, it is thus possible to stabilize three different domains, each with the   principle axis of the EFG, $V_{zz}$, pointing along any of the 3 equivalent crystal axes.  Further, local magnetic field has to be parallel to $V_{\rm zz}$ in each domain. The facts that the splitting is the largest for $H \| [001]$  (\mbox{Fig. 2B}), 
 and that only 3 peaks per set are observed for ${\mathbf H} \| [110]$ imply that two domains are plausible   in the crystal.  One domain  is characterized by  pure uniaxial $3z^2 - r^2$ distortions where $V_{\rm zz}$   is in [001] direction, while   the other is distinguished by $x^2 - y^2$ distortions where $V_{\rm zz}$ is then in the (110) plane.  
In the simplest case $V_{\rm zz}$ is parallel  to [001] direction with $\eta = 0$ indicating tetragonal local symmetry. In the second case $V_{\rm zz}$ is aligned along [100] direction with  $\eta$ is of the order of 1,   implying orthorhombic local symmetry.  
To determine the exact local symmetry, splittings $\delta_{\rm q}$ obtained for ${\mathbf H}$ rotated about one of the crystalline axis and about  [110] direction have to be analyzed.

         {\bf Calculation of Internal Fields:}
   In the LRO phase, the component of the internal hyperfine field parallel to $H$, at an  Na   site,
is given by $H^{\|}_{\rm int}  = \hat {\mathbf h} \cdot \sum_{\left<i\right>} \mathbb{A}_{\rm i}\cdot \boldsymbol{\mu}_{\rm i} ,$
where   $\hat {\mathbf h}$ is a unit vector in the applied field direction,  $\mathbb A_{\rm i}$ is the symmetric $3 \times 3$ hyperfine coupling tensor with  the i$^{th}$ nearest-neighbor Os  atom and $\boldsymbol{\mu}_{\rm i}$ is
its magnetic moment (see Supplementary Note 4). In the PM phase, hyperfine coupling tensor  is diagonal. 
Due to complexity   of the orbitals mediating the exchange paths, which can induce multipolar exchange interactions \cite{KrempaRev14,ChenBalents10} between neighboring Os spins and broken cubic symmetry, the off-diagonal elements of the hyperfine tensor $A_{ij}$ are nonzero in the LRO phase. 
We point out that even if moment is not exclusively localized on Os site but the spin density is distributed to O \cite{Pickett15}, our modeling of $H^{\rm i}_{\rm int}$ 
is valid. This is  because the complexity of the spin density is  accounted for in  $\mathbb{A}$. For simplicity, we treat moment as $S=1/2$ localized on Os as was done in \cite{Balents14}. 
By performing full lattice sum,  we calculate the  local \hu and \hst at the Na site as well as Na NMR spectra, which is a histogram of the local field component projected along the applied field, as the direction of the applied field is rotated in   $(1 \bar1 0)$ plane of the crystal. 
    \\
  


\vspace{0.5cm}

{\bf Figure Legends:}  
\vspace{0.2cm}

{\bf 1.}     {\bf   Phase diagram of  \BaOsS deduced from the  NMR spectra.} {(\bf a}) Temperature evolution of $^{23}$Na spectra  
at 9 T  (and at 15 T, shaded trace) magnetic field   applied parallel to [001] crystalline axis.   
 Narrow single peak spectra characterize high temperature paramagnetic (PM) state.  At intermediate temperatures, broader and more complex spectra reveal the appearance of electric field gradient (EFG) induced by breaking  of local cubic symmetry. 
Splitting into 2 sets of triplet lines (labeled as I and II),   reflecting the existence of two distinct magnetic sites in the lattice, is evident at lower temperatures. 
Zero of frequency is defined as $\omega_{0} =  \, ^{23}\gamma \,H$.  Abbreviation: PM, paramagnetic; BLPS, broken local cubic symmetry; and, cFM,  canted ferromagnetic.
 The charge density in the theoretically predicted quadrupolar phase \cite{KrempaRev14} is sketched in the inset.  ({\bf b})   Sketch of the phase diagram based on our NMR measurements. Squares indicate onset  temperature   for the local cubic symmetry breaking,     determined from our NMR data as explained in the text, in the PM  phase. Circles denote $T_{\rm c}$, transition temperature into canted ferromagnetic (cFM) phase, as deduced from the NMR data, while diamond marks $T_c$ as determined from thermodynamic measurements in \mbox{Ref. \cite{Erickson07}}.
The solid line  indicates phase transition into cFM state and also 
possible tetragonal-to-orthorhombic  phase transition. The dashed  line denotes cross over to the broken local point symmetry (BLPS) phase, as detected by NMR.
 ( {\bf c}) The high-temperature undistorted crystal structure of \BaOs. In this case, point symmetry at the Na site is cubic leading to zero electric field gradient (EFG). Principal crystallographic axes are   shown as well. \\

{\bf 2.} {\bf Local cubic symmetry breaking in the ordered phase.} ({\bf a}) $^{23}$Na spectra    in low temperature ordered state as a function of the angle between the applied magnetic field and [001] crystalline axis. In this case, ${\mathbf H}$ was rotated in the $(1 \bar1 0)$ plane of the crystal which contains three high symmetry directions: [001], [111], and [110]. 
({\bf b}) The mean peak-to-peak splitting $(\delta_{\rm q})$ between any two adjacent peaks of the triplets I and II. 
Error bars reflect the scattering of deduced $(\delta_{\rm q})$ values. 
Solid line is the fit to 
$|(3   \cos^{2} \theta -1)/2 |$,  where $\theta$ denotes the angle between the principal axis of the EFG $(V_{\rm zz})$ and the applied magnetic field $({\mathbf H})$ as depicted for nuclei spins $I = 3/2$ with magnetic moment ${\boldsymbol{ \mu}}$   (see Supplementary Note 5).    
({\bf c}) Schematic of a proposed lattice distortion involving the O$^{{2-}}$ ions.   The distortion  breaks cubic  symmetry at the Na site giving rise to 
finite electric field gradient (EFG). Different types   and colors of double  arrows illustrate unequal magnitude of oxygen displacements.  
Displacements along different axes can comprise either from compression or elongation of the oxygen bonds (see Supplementary Note 5).    Depicted distortion magnitude is not to scale with the distance between  nearest neighbor Os atoms, and is amplified for clarity. ({\bf d}) Schematic of the energy levels of a spin-3/2 nucleus in a finite magnetic field and in the presence of quadrupolar interaction with EFG, generated by surrounding electronic charges, and resulting NMR spectra. 
In the absence of quadrupolar interaction spectrum consists of a single narrow line at frequency $\omega$ and of width $\Delta_{\omega}$. 
In the presence of quadrupolar interaction the centre transition remains at frequency  $\omega$, while the satellite transitions appear at frequencies    shift by $\pm$ \dq, proportional to the magnitude of the EFG. For small values of the EFG, satellite transition cannot be resolved and only line broadening is observed. 
Strictly speaking, there is also a broadening due to the distribution of magnitude of the EFG  itself, but this is manifested only on the satellites and not on the central transition. In our case, this distribution can be neglected as all the lines show the same width.\\

{\bf 3.} {\bf Temperature dependence of the uniform and staggered fields.}  Absolute value of the uniform internal field, $H_{\rm u}$, ({\bf a})  and the staggered internal field, $H_{{\rm stag}}$,  ({\bf b}) as a function of reduced temperature for various ${\mathbf H} \| [001]$. Typical error bars are on the order of a few per cent and not shown for clarity. 
Lines are guide to the eyes. \\

{\bf 4.} {\bf  Resolution of the spin orientation in the ordered phase.} ({\bf a})  The magnitude of the  
internal field associated with triplet I $(H_{\rm I})$ and II $(H_{\rm II})$ (solid symbols), and the first moment of the entire spectrum (uniform field, $H_{\rm u}$) as a function of the angle between the applied magnetic field and [001] crystalline axis at 8 K and 15 T. Open symbols depict the angular dependence of  the staggered   field, $H_{{\rm stag}}$.  Typical error bars are on the order of a few per cent and not shown for clarity. 
 Dotted lines are guide to the eyes. Solid lines are calculated fields from the spin model sketched in part D, as described in the text. 
({\bf b})  Schematic of   the net magnetization   in the XY plane in   the spin model, consistent with our data and proposed in \cite{Balents14}.  Arrows of different shades  depict spins from two sub-lattices,  labeled as A and B. These spins are canted by angle $\pm \phi$ with respect to [110] direction. ({\bf c})  Comparison of the angular dependence of bulk magnetization per FU (formula unit) from \mbox{Ref. \cite{Erickson07}} (open symbols) and $H_{\rm u}$ determined from our NMR measurements. ({\bf d})  Schematic of the spin model consistent with our data and proposed in \cite{Balents14}. Different shades and orientation of arrows indicate distinct ionic (spin) environments on Os site.  
 Planes containing moments from sub-lattice A and B are shown.
Lines are guide to the eyes. \\

\noindent {\bf Acknowledgements}
 We would like to thank  L. Balents and M. Horvati{\'c}   for   illuminating discussions.
    The study was supported in part by the the National Science Foundation (DMR-0547938 and DMR-1608760).       The study at the NHMFL was supported by the National Science Foundation under Cooperative Agreement no. DMR95-27035 and the State of Florida. 
   Work at Stanford University was  supported by the DOE, Office of Basic Energy Sciences, under Contract No. 
   DE-AC02-76SF00515. \\

\vspace{0.0cm}
\end{document}